\documentstyle[twocolumn,aps]{revtex}

\begin{document}
\title
{\bf Rotating Bose-Einstein Condensates with a large number of Vortices.
}
\author{V. Subrahmanyam}
\address{
Department of Physics, Indian Institute of Technology, Kanpur-208016, India.\\
\centerline{\rm 1 June 2002}
\parbox[t]{14 truecm}{\small
We study a harmonically-confined Bose-Einstein condensate under rotation.
Vortex lattice configurations are investigated through a variational approach.
Vortices with
more than a unit of angular momentum are not stable.
We explicitly show that the critical rotational
frequency is quite small, as was reported in the MIT experiment. The
width of the condensate in the xy plane and 
along z, and the vortex density are determined as functions of the 
rotational frequency. The aspect ratio becomes small for large rotational
frequencies, making the condensate an effectively two-dimensional system. 
}}
\maketitle
\narrowtext 
 
In recent experiments\cite{ref1a,ref1b}, a large number of vortices (about 150) have
been generated in a Bose-Einstein condensate of Na atoms. The vortices form
a hexagonal lattice,
in a quasi equilibrium, showing a variety of properties 
which differ from a equilibrium system. In these experiments, a large 
angular momentum is imparted to the condensate by rotating it close to the
quadrupolar frequency. In addition, the condensate is subjected to an asymmetric
harmonic confining potential: $\omega_z/2\pi \sim 20 Hz$, $\omega_{xy}/2\pi
\sim 80 Hz$, the rotation axis being along z in the MIT experiment\cite{ref1a}.
The high angular momentum states of the Bose condensate are similar to the
quantum Hall systems and type-II superconductors. 
If the rotational frequency $\Omega$ is exactly same
as in-plane harmonic frequency $\omega_{xy}$, these states can be mapped
to the lowest Landau level of quantum Hall systems\cite{ref2}. 

The angular momentum that is transfered to the condensate is carried by 
vortices, each vortex carrying a quantum of angular momentum, similar to
the vortices in a superfluid. The vortices have an effective repulsion, which
we shall see later from a variational calculation. This 
implies that a vortex with two quanta of angular momentum is unstable against
separating into two unit vortices. When a large angular momentum is 
deposited on the condensate, the vortices form a regular lattice in the
xy plane. The unit cell area $\nu$ of the vortex lattice will depend on
the rotational frequency, the harmonic frequencies, and the two-particle
scattering length of Na atoms, through the parameters charactering the 
condensate wavefunction. In the previous work of Ho\cite{ref2}, the energy of
the vortex lattice has been calculated through a variational approach within
the Gross-Pitaevskii paradigm, using a Thomas-Fermi approximation.
The calculated critical rotational frequency does not agree with the
MIT experiment\cite{ref2a}. We will carry out
a variational calculation, with a variational condensate wavefunction that 
is similar to the one used in\cite{ref2}, that allows us to estimate the
critical rotational frequency which is in excellent agreement with the
experiment, and the aspect ratio of the condensate (the
ration of the rms widths in the xy plane and along the z-direction).

The condensate wavefunction is determined from minimizing the Gross-Pitaevskii
functional\cite{ref2a}
\begin{equation}
K =\int \psi^{\star}[H_{xy}-\Omega L_z +H_z]\psi d^3r +{gN\over2}\int 
|\psi|^4 d^3r
\end{equation}
where the single-particle Hamiltonians along $z$ and in the $xy$ plane are
$H_z={-\hbar^2\over 2M}\nabla_z^2+{1\over2M}\omega_z^2z^2,H_{xy}=-{\hbar^2\over 
2M}\nabla_{xy}^2+{1\over2M}\omega_{xy}^2(x^2+y^2)$; $L_z$ is the angular momentum
along $z$ direction along which the condensate is rotated with a frequency
$\Omega$; the two-particle interaction strength is $g=4\pi\hbar^2 a_{sc}/M$ 
in terms of the s-wave scattering length; $N$ is the number of atoms in the
condensate, and $M$ is the mass of the Na atom. The wavefunction $\psi$ is
normalized to unity. The Hamiltonian
in the xy plane is similar to that of a charged particle in a magnetic field if
$eB/Mc=2\omega_{xy}$, at $\omega_{xy}=\Omega$. In this case the energy levels
are Landau levels, with the lowest level degenerate for angular momentum
quantum number $m=0,1,2..$ The degeneracy gets lifted for $\omega_{xy}\ne 
\Omega.$

In the absence of interactions, for $g=0$, the lowest energy state is a direct
product of harmonic oscillator wavefunctions along z direction and in the
xy plane, with uniform densities (modulo the gaussian variation). However,
for $g\ne 0$, states with nonuniform densities, more specifically with
density vanishing at points signifying vortices, are favorable. These states
will become stable for $\Omega>\Omega_c$, where $\Omega_c$ is the critical
rotational frequency. We will follow in spirit\cite{ref3}
to estimate the critical rotational frequency.
Let us write $\psi(\vec r,z)=\Phi_m(\vec r)\chi(z)$,by separating the z 
direction and $xy$ plane, where $\chi(z)=c_z
\exp(-z^2\gamma^2/ 2a_z^2)$, $\Phi_m(\vec r)=c_m r^m \exp(im\phi)
exp(-r^2\beta^2/2a_{xy}^2)$; $a_z=\sqrt{\hbar/M\omega_z}$ and $a_{xy}=\sqrt{\hbar
/M\omega_{xy}}$ are the oscillator lengths,
$\gamma,\beta$ are variational parameters, and $c_m,c_z$ are normalization
constants.
The 
wavefunction carries an angular momentum $m$, and since $|\Phi|^2\sim r^{2m}$,
it also signifies a vortex at the origin of strength $m$. For $\Omega$=0, the
uniform density state (modulo a gaussian variation), viz. $m=0$ has the lowest 
energy
for $\gamma=1,\beta=1$ in the absence of the interactions, corresponding to
the groundstate of the harmonic oscillator.  For $g\ne 0$, the
parameters $\beta,\gamma$ will change substantially. The energy of the
configuration $\Phi_m$ is (in units of the oscillator energy $\hbar 
\omega_{xy}$)
\begin{eqnarray}
K_m= \lbrace {1\over 2}({1\over\beta^2}+\beta^2)-{\Omega\over \omega_{xy}}
\rbrace \Gamma_1
+{\omega_z \over 4\omega_{xy}}({1\over\gamma^2}+\gamma^2) \nonumber\\
+G\gamma\beta^2\Gamma_2+{\Omega\over \omega_{xy}}
\end{eqnarray}
where the dimensionless parameter $G= Na_{sc}/a_z\sqrt{2\pi}$ 
(using experimental values
$N=5 \times 10^7,a_{sc}= 50$Bohr radii,$a_z=0.9\times  10^5$Bohr
radii, the interaction strength is $G\sim 1.1\times 10^4$). The coefficient of 
the first term in the above equation, $\Gamma_1=<r^2>\beta^2/a_{xy}^2$ is just 
the rms width of 
the wavefunction in units of $a_{xy}^2/\beta^2; \Gamma_1(m)=m+1$. 
The coefficient of the
last term, $\Gamma_2=2\pi a_{xy}^2/\beta^2 \int|\Phi|^4$, is the interaction 
contribution; $\Gamma_2(m)=2m!/m!m!2^m$. 
Minimizing the energy w.r.t. parameters $\gamma$ and $\beta$ for the cases of
$m=0$ (no vortex configuration) and $m=1$ (one vortex of unit strength),
we obtain $\gamma_0=\gamma_1= (\omega_z/\omega_{xy})^{2/5}(2G)^{-1/5},
\beta_0^2=\gamma_0^2(\omega_{xy}/\omega_z),\beta_1^2=2\beta_0^2$. For the
experimental values of $\omega_z/\omega_{xy}=1/4,G=10^4$ we get $\gamma_0=
0.08,\beta_0=0.16$ (in contrast, $\gamma=\beta=1$, when $g=0$ for the no-vortex
state). And the
energy difference between the two configurations
$$
K_1-K_0={3\gamma_0^2\omega_{xy}\over \omega_z}-
{\Omega\over \omega_{xy}}\equiv{{\Omega_c-\Omega}\over\omega_{xy}}
$$
will be negative for $\Omega>\Omega_c$ where the critical rotational frequency
is ${\Omega_c/\omega_{xy}}\equiv{3\gamma_0^2\omega_{xy}/\omega_z}$
is about 0.077 for the experimental values of $G=10^4,\omega_z/\omega_{xy}=1/4$.
This compares very well with the MIT experiments\cite{ref1a}, where vortices
have been observed for frequencies as low as 7 Hz, with the harmonic oscillator
frequency of 80 Hz. 

The above discussion establishes that for $\Omega>\Omega_c$ vortex 
configurations are stable against a uniform density state. We examined 
configurations with larger strength of vortex, viz. for $m$=2,3.. 
A larger $m$
will lower the interaction term, which is clear from the coefficient, but
will increase the rms width of the wavefunction. 
For $m=2$, the energy is minimized for $\gamma_2=\gamma_0(8/9)^{1/5}$, and
$\beta_2^2=2\beta_0^26^{1/5}$, implying $K_2-K_1=1.25
\gamma_0^2((9/8)^{2/5}-1)\omega_z/\omega_{xy} >0$. It can be seen that for $m> 2$, the energy is
not favorable, implying a vortex with angular momentum larger than one unit
is not stable. The $m=2$ state is unstable against a state with two $m=1$
vortices separated by some distance. The energy is lower for a larger 
separation, implying an effective repulsion between the vortices.
We can investigate states with a large angular momentum
by looking at configurations with a large number of vortices with unit
strength. This is tantamount to separating the vortices in state $\Phi_m$.
Instead of a m-vortex sitting at the origin, we will have m one-vortices 
spread all over.
The effective repulsion between the vortices gives a finite
spacing. A regular lattice structure is easily probed, either triangular or
rectangular (the only Bravais lattices in two dimensions).

Let us consider a trial wavefunction\cite{ref2}, with $q$ vortices each
carrying a unit of angular momentum
\begin{equation}
\Phi=c_q e^{-r^2\beta^2/2a_{xy}^2}
\prod_{l=1}^q (z-z_l)=\sum_{m=0}^q x_m \Phi_m
\end{equation}
where we introduce a vortex function $f(z)=\prod (z-z_l), z_l=
x_l+iy_l$ is the complex coordinate of the l'th vortex (carrying a unit
of angular momentum). We have
also written the wavefunction as a superposition of various configurations 
$\Phi_m$ (carrying one vortex of strength $m$). Noting that $\int \Phi_m H_{xy}
\Phi_{m^\prime}=0$, for $m\ne m^\prime$,  the energy of this state
can be 
written as before (Eq.2) 
with
\begin{equation}
\Gamma_1={\beta^2\over a_{xy}^2} {\int |\Phi|^2 r^2 d^2r \over \int |\Phi|^2 d^2r}
\end{equation} 
\noindent and  
\begin{equation}
\Gamma_2= {2\pi a_{xy}^2\over \beta^2}\int |\Phi|^4 d^2r.
\end{equation}
The evaluation of the above coefficients
$\Gamma_1, \Gamma_2$, is quite involved, as the computation of the integrals
is not straightforward, because of the presence of the vortex function
$f(z)$. We shall try to extract the leading order behaviour of these
coefficients as functions of the ratio $\pi a_{xy}^2/\beta^2\nu$, where $\nu$ is
the unit cell area of the vortex lattice.

Firstly, we note that the vortex function can be written as $f(z)=z(z^6-b_1^6)
(z^6-b_2^6)(z^6-b_3^6)(z^{12}-b_4^{12})..$ , for
a triangular lattice (actually a hexagonal lattice, including the z direction)
where $b_n$ is the nth neighbor distance in units of $a_{xy}/\beta$. 
Each $b_n$ occurs with a power which is equal to the coordination number of the
n'th neighbor vortex. For triangular lattice the coordination is either 6 or
12. In the case of a square
lattice the coordination number is either 4 or 8.
Here we have fixed the origin on
one of the vortices. In the product $(z-z_1)..(z-z_6)$ of the first neighbor
vortices, only two terms are nonzero, viz. $z^6, z_1z_2..z_6=-b_1^6$. 
Similarly for the other neighbour vortices.   We now resort to a simple
approximation to estimate $f(z)$. Let us write
\begin{eqnarray}
\log f(z)&=&\sum \log (z-z_i) \nonumber \\
&=& {1\over \nu}\int_0^R r^\prime dr^\prime \int_0^{2\pi}d\theta
\log{(z-r^\prime e^{i\theta})},\end{eqnarray}
where we replaced the sum by an integral, and $\nu$ is the unit cell volume
of the vortex lattice. In the above approximation, namely replacing the
density of vortices by a constant, we have ignored the finite-size 
corrections that arise, as the lattice size is not too big (about 150 vortices
or less). This approximation is similar to the average-vortex approximation
of \cite{ref2}.
The details of finite-size effects on the vortex function will be
given else where. Here, we will keep only the smooth part of the density of 
vortices. In the integral the upper limit, viz.
the size of the vortex lattice, is taken to be $R$, with $\pi R^2/\nu=q$, the
total number of vortices.
The integral is easily done, 
by doing an integration by parts in 
$r^\prime$ variable, keeping 
the log function as the second function, we get (using $|z|=r$)
\begin{equation}
\log f\approx iq\pi+\log({q\nu\over \pi})^{q/2} + {\pi r^2\over 2\nu}.
\end{equation}
Now, $|f(z)|^2=(q\nu/ \pi)^q \exp{(\pi r^2/\nu})$, and in turn the 
condensate density is given by

\setlength{\unitlength}{0.240900pt}
\ifx\plotpoint\undefined\newsavebox{\plotpoint}\fi
\begin{picture}(1005,765)(0,0)
\font\gnuplot=cmr10 at 10pt
\gnuplot
\sbox{\plotpoint}{\rule[-0.200pt]{0.400pt}{0.400pt}}%
\put(161.0,158.0){\rule[-0.200pt]{4.818pt}{0.400pt}}
\put(141,158){\makebox(0,0)[r]{420}}
\put(924.0,158.0){\rule[-0.200pt]{4.818pt}{0.400pt}}
\put(161.0,311.0){\rule[-0.200pt]{4.818pt}{0.400pt}}
\put(141,311){\makebox(0,0)[r]{455}}
\put(924.0,311.0){\rule[-0.200pt]{4.818pt}{0.400pt}}
\put(161.0,463.0){\rule[-0.200pt]{4.818pt}{0.400pt}}
\put(141,463){\makebox(0,0)[r]{490}}
\put(924.0,463.0){\rule[-0.200pt]{4.818pt}{0.400pt}}
\put(161.0,616.0){\rule[-0.200pt]{4.818pt}{0.400pt}}
\put(141,616){\makebox(0,0)[r]{525}}
\put(924.0,616.0){\rule[-0.200pt]{4.818pt}{0.400pt}}
\put(371.0,123.0){\rule[-0.200pt]{0.400pt}{4.818pt}}
\put(371,82){\makebox(0,0){2}}
\put(371.0,705.0){\rule[-0.200pt]{0.400pt}{4.818pt}}
\put(650.0,123.0){\rule[-0.200pt]{0.400pt}{4.818pt}}
\put(650,82){\makebox(0,0){4}}
\put(650.0,705.0){\rule[-0.200pt]{0.400pt}{4.818pt}}
\put(930.0,123.0){\rule[-0.200pt]{0.400pt}{4.818pt}}
\put(930,82){\makebox(0,0){6}}
\put(930.0,705.0){\rule[-0.200pt]{0.400pt}{4.818pt}}
\put(161.0,123.0){\rule[-0.200pt]{188.625pt}{0.400pt}}
\put(944.0,123.0){\rule[-0.200pt]{0.400pt}{145.022pt}}
\put(161.0,725.0){\rule[-0.200pt]{188.625pt}{0.400pt}}
\put(40,424){\makebox(0,0){$\log{|f|^2}$}}
\put(552,21){\makebox(0,0){$\pi r^2/\nu$}}
\put(161.0,123.0){\rule[-0.200pt]{0.400pt}{145.022pt}}
\put(491,594){\makebox(0,0)[r]{From Eq.7}}
\put(511.0,594.0){\rule[-0.200pt]{24.090pt}{0.400pt}}
\put(161,131){\usebox{\plotpoint}}
\put(161,130.67){\rule{1.927pt}{0.400pt}}
\multiput(161.00,130.17)(4.000,1.000){2}{\rule{0.964pt}{0.400pt}}
\put(169,131.67){\rule{1.927pt}{0.400pt}}
\multiput(169.00,131.17)(4.000,1.000){2}{\rule{0.964pt}{0.400pt}}
\put(177,132.67){\rule{1.927pt}{0.400pt}}
\multiput(177.00,132.17)(4.000,1.000){2}{\rule{0.964pt}{0.400pt}}
\put(185,134.17){\rule{1.700pt}{0.400pt}}
\multiput(185.00,133.17)(4.472,2.000){2}{\rule{0.850pt}{0.400pt}}
\put(193,135.67){\rule{1.927pt}{0.400pt}}
\multiput(193.00,135.17)(4.000,1.000){2}{\rule{0.964pt}{0.400pt}}
\put(201,137.17){\rule{1.500pt}{0.400pt}}
\multiput(201.00,136.17)(3.887,2.000){2}{\rule{0.750pt}{0.400pt}}
\put(208,138.67){\rule{1.927pt}{0.400pt}}
\multiput(208.00,138.17)(4.000,1.000){2}{\rule{0.964pt}{0.400pt}}
\put(216,140.17){\rule{1.700pt}{0.400pt}}
\multiput(216.00,139.17)(4.472,2.000){2}{\rule{0.850pt}{0.400pt}}
\put(224,141.67){\rule{1.927pt}{0.400pt}}
\multiput(224.00,141.17)(4.000,1.000){2}{\rule{0.964pt}{0.400pt}}
\put(232,143.17){\rule{1.700pt}{0.400pt}}
\multiput(232.00,142.17)(4.472,2.000){2}{\rule{0.850pt}{0.400pt}}
\put(240,145.17){\rule{1.700pt}{0.400pt}}
\multiput(240.00,144.17)(4.472,2.000){2}{\rule{0.850pt}{0.400pt}}
\put(248,147.17){\rule{1.700pt}{0.400pt}}
\multiput(248.00,146.17)(4.472,2.000){2}{\rule{0.850pt}{0.400pt}}
\multiput(256.00,149.61)(1.579,0.447){3}{\rule{1.167pt}{0.108pt}}
\multiput(256.00,148.17)(5.579,3.000){2}{\rule{0.583pt}{0.400pt}}
\put(264,152.17){\rule{1.700pt}{0.400pt}}
\multiput(264.00,151.17)(4.472,2.000){2}{\rule{0.850pt}{0.400pt}}
\put(272,154.17){\rule{1.700pt}{0.400pt}}
\multiput(272.00,153.17)(4.472,2.000){2}{\rule{0.850pt}{0.400pt}}
\multiput(280.00,156.61)(1.579,0.447){3}{\rule{1.167pt}{0.108pt}}
\multiput(280.00,155.17)(5.579,3.000){2}{\rule{0.583pt}{0.400pt}}
\put(288,159.17){\rule{1.500pt}{0.400pt}}
\multiput(288.00,158.17)(3.887,2.000){2}{\rule{0.750pt}{0.400pt}}
\multiput(295.00,161.61)(1.579,0.447){3}{\rule{1.167pt}{0.108pt}}
\multiput(295.00,160.17)(5.579,3.000){2}{\rule{0.583pt}{0.400pt}}
\multiput(303.00,164.61)(1.579,0.447){3}{\rule{1.167pt}{0.108pt}}
\multiput(303.00,163.17)(5.579,3.000){2}{\rule{0.583pt}{0.400pt}}
\put(311,167.17){\rule{1.700pt}{0.400pt}}
\multiput(311.00,166.17)(4.472,2.000){2}{\rule{0.850pt}{0.400pt}}
\multiput(319.00,169.61)(1.579,0.447){3}{\rule{1.167pt}{0.108pt}}
\multiput(319.00,168.17)(5.579,3.000){2}{\rule{0.583pt}{0.400pt}}
\multiput(327.00,172.60)(1.066,0.468){5}{\rule{0.900pt}{0.113pt}}
\multiput(327.00,171.17)(6.132,4.000){2}{\rule{0.450pt}{0.400pt}}
\multiput(335.00,176.61)(1.579,0.447){3}{\rule{1.167pt}{0.108pt}}
\multiput(335.00,175.17)(5.579,3.000){2}{\rule{0.583pt}{0.400pt}}
\multiput(343.00,179.61)(1.579,0.447){3}{\rule{1.167pt}{0.108pt}}
\multiput(343.00,178.17)(5.579,3.000){2}{\rule{0.583pt}{0.400pt}}
\multiput(351.00,182.61)(1.579,0.447){3}{\rule{1.167pt}{0.108pt}}
\multiput(351.00,181.17)(5.579,3.000){2}{\rule{0.583pt}{0.400pt}}
\multiput(359.00,185.60)(1.066,0.468){5}{\rule{0.900pt}{0.113pt}}
\multiput(359.00,184.17)(6.132,4.000){2}{\rule{0.450pt}{0.400pt}}
\multiput(367.00,189.61)(1.579,0.447){3}{\rule{1.167pt}{0.108pt}}
\multiput(367.00,188.17)(5.579,3.000){2}{\rule{0.583pt}{0.400pt}}
\multiput(375.00,192.60)(0.920,0.468){5}{\rule{0.800pt}{0.113pt}}
\multiput(375.00,191.17)(5.340,4.000){2}{\rule{0.400pt}{0.400pt}}
\multiput(382.00,196.60)(1.066,0.468){5}{\rule{0.900pt}{0.113pt}}
\multiput(382.00,195.17)(6.132,4.000){2}{\rule{0.450pt}{0.400pt}}
\multiput(390.00,200.60)(1.066,0.468){5}{\rule{0.900pt}{0.113pt}}
\multiput(390.00,199.17)(6.132,4.000){2}{\rule{0.450pt}{0.400pt}}
\multiput(398.00,204.60)(1.066,0.468){5}{\rule{0.900pt}{0.113pt}}
\multiput(398.00,203.17)(6.132,4.000){2}{\rule{0.450pt}{0.400pt}}
\multiput(406.00,208.60)(1.066,0.468){5}{\rule{0.900pt}{0.113pt}}
\multiput(406.00,207.17)(6.132,4.000){2}{\rule{0.450pt}{0.400pt}}
\multiput(414.00,212.60)(1.066,0.468){5}{\rule{0.900pt}{0.113pt}}
\multiput(414.00,211.17)(6.132,4.000){2}{\rule{0.450pt}{0.400pt}}
\multiput(422.00,216.60)(1.066,0.468){5}{\rule{0.900pt}{0.113pt}}
\multiput(422.00,215.17)(6.132,4.000){2}{\rule{0.450pt}{0.400pt}}
\multiput(430.00,220.59)(0.821,0.477){7}{\rule{0.740pt}{0.115pt}}
\multiput(430.00,219.17)(6.464,5.000){2}{\rule{0.370pt}{0.400pt}}
\multiput(438.00,225.60)(1.066,0.468){5}{\rule{0.900pt}{0.113pt}}
\multiput(438.00,224.17)(6.132,4.000){2}{\rule{0.450pt}{0.400pt}}
\multiput(446.00,229.59)(0.821,0.477){7}{\rule{0.740pt}{0.115pt}}
\multiput(446.00,228.17)(6.464,5.000){2}{\rule{0.370pt}{0.400pt}}
\multiput(454.00,234.60)(1.066,0.468){5}{\rule{0.900pt}{0.113pt}}
\multiput(454.00,233.17)(6.132,4.000){2}{\rule{0.450pt}{0.400pt}}
\multiput(462.00,238.59)(0.710,0.477){7}{\rule{0.660pt}{0.115pt}}
\multiput(462.00,237.17)(5.630,5.000){2}{\rule{0.330pt}{0.400pt}}
\multiput(469.00,243.59)(0.821,0.477){7}{\rule{0.740pt}{0.115pt}}
\multiput(469.00,242.17)(6.464,5.000){2}{\rule{0.370pt}{0.400pt}}
\multiput(477.00,248.59)(0.821,0.477){7}{\rule{0.740pt}{0.115pt}}
\multiput(477.00,247.17)(6.464,5.000){2}{\rule{0.370pt}{0.400pt}}
\multiput(485.00,253.59)(0.821,0.477){7}{\rule{0.740pt}{0.115pt}}
\multiput(485.00,252.17)(6.464,5.000){2}{\rule{0.370pt}{0.400pt}}
\multiput(493.00,258.59)(0.821,0.477){7}{\rule{0.740pt}{0.115pt}}
\multiput(493.00,257.17)(6.464,5.000){2}{\rule{0.370pt}{0.400pt}}
\multiput(501.00,263.59)(0.671,0.482){9}{\rule{0.633pt}{0.116pt}}
\multiput(501.00,262.17)(6.685,6.000){2}{\rule{0.317pt}{0.400pt}}
\multiput(509.00,269.59)(0.821,0.477){7}{\rule{0.740pt}{0.115pt}}
\multiput(509.00,268.17)(6.464,5.000){2}{\rule{0.370pt}{0.400pt}}
\multiput(517.00,274.59)(0.671,0.482){9}{\rule{0.633pt}{0.116pt}}
\multiput(517.00,273.17)(6.685,6.000){2}{\rule{0.317pt}{0.400pt}}
\multiput(525.00,280.59)(0.821,0.477){7}{\rule{0.740pt}{0.115pt}}
\multiput(525.00,279.17)(6.464,5.000){2}{\rule{0.370pt}{0.400pt}}
\multiput(533.00,285.59)(0.671,0.482){9}{\rule{0.633pt}{0.116pt}}
\multiput(533.00,284.17)(6.685,6.000){2}{\rule{0.317pt}{0.400pt}}
\multiput(541.00,291.59)(0.671,0.482){9}{\rule{0.633pt}{0.116pt}}
\multiput(541.00,290.17)(6.685,6.000){2}{\rule{0.317pt}{0.400pt}}
\multiput(549.00,297.59)(0.581,0.482){9}{\rule{0.567pt}{0.116pt}}
\multiput(549.00,296.17)(5.824,6.000){2}{\rule{0.283pt}{0.400pt}}
\multiput(556.00,303.59)(0.671,0.482){9}{\rule{0.633pt}{0.116pt}}
\multiput(556.00,302.17)(6.685,6.000){2}{\rule{0.317pt}{0.400pt}}
\multiput(564.00,309.59)(0.671,0.482){9}{\rule{0.633pt}{0.116pt}}
\multiput(564.00,308.17)(6.685,6.000){2}{\rule{0.317pt}{0.400pt}}
\multiput(572.00,315.59)(0.671,0.482){9}{\rule{0.633pt}{0.116pt}}
\multiput(572.00,314.17)(6.685,6.000){2}{\rule{0.317pt}{0.400pt}}
\multiput(580.00,321.59)(0.671,0.482){9}{\rule{0.633pt}{0.116pt}}
\multiput(580.00,320.17)(6.685,6.000){2}{\rule{0.317pt}{0.400pt}}
\multiput(588.00,327.59)(0.569,0.485){11}{\rule{0.557pt}{0.117pt}}
\multiput(588.00,326.17)(6.844,7.000){2}{\rule{0.279pt}{0.400pt}}
\multiput(596.00,334.59)(0.671,0.482){9}{\rule{0.633pt}{0.116pt}}
\multiput(596.00,333.17)(6.685,6.000){2}{\rule{0.317pt}{0.400pt}}
\multiput(604.00,340.59)(0.569,0.485){11}{\rule{0.557pt}{0.117pt}}
\multiput(604.00,339.17)(6.844,7.000){2}{\rule{0.279pt}{0.400pt}}
\multiput(612.00,347.59)(0.569,0.485){11}{\rule{0.557pt}{0.117pt}}
\multiput(612.00,346.17)(6.844,7.000){2}{\rule{0.279pt}{0.400pt}}
\multiput(620.00,354.59)(0.671,0.482){9}{\rule{0.633pt}{0.116pt}}
\multiput(620.00,353.17)(6.685,6.000){2}{\rule{0.317pt}{0.400pt}}
\multiput(628.00,360.59)(0.569,0.485){11}{\rule{0.557pt}{0.117pt}}
\multiput(628.00,359.17)(6.844,7.000){2}{\rule{0.279pt}{0.400pt}}
\multiput(636.00,367.59)(0.492,0.485){11}{\rule{0.500pt}{0.117pt}}
\multiput(636.00,366.17)(5.962,7.000){2}{\rule{0.250pt}{0.400pt}}
\multiput(643.00,374.59)(0.569,0.485){11}{\rule{0.557pt}{0.117pt}}
\multiput(643.00,373.17)(6.844,7.000){2}{\rule{0.279pt}{0.400pt}}
\multiput(651.00,381.59)(0.494,0.488){13}{\rule{0.500pt}{0.117pt}}
\multiput(651.00,380.17)(6.962,8.000){2}{\rule{0.250pt}{0.400pt}}
\multiput(659.00,389.59)(0.569,0.485){11}{\rule{0.557pt}{0.117pt}}
\multiput(659.00,388.17)(6.844,7.000){2}{\rule{0.279pt}{0.400pt}}
\multiput(667.00,396.59)(0.569,0.485){11}{\rule{0.557pt}{0.117pt}}
\multiput(667.00,395.17)(6.844,7.000){2}{\rule{0.279pt}{0.400pt}}
\multiput(675.00,403.59)(0.494,0.488){13}{\rule{0.500pt}{0.117pt}}
\multiput(675.00,402.17)(6.962,8.000){2}{\rule{0.250pt}{0.400pt}}
\multiput(683.00,411.59)(0.494,0.488){13}{\rule{0.500pt}{0.117pt}}
\multiput(683.00,410.17)(6.962,8.000){2}{\rule{0.250pt}{0.400pt}}
\multiput(691.00,419.59)(0.569,0.485){11}{\rule{0.557pt}{0.117pt}}
\multiput(691.00,418.17)(6.844,7.000){2}{\rule{0.279pt}{0.400pt}}
\multiput(699.00,426.59)(0.494,0.488){13}{\rule{0.500pt}{0.117pt}}
\multiput(699.00,425.17)(6.962,8.000){2}{\rule{0.250pt}{0.400pt}}
\multiput(707.00,434.59)(0.494,0.488){13}{\rule{0.500pt}{0.117pt}}
\multiput(707.00,433.17)(6.962,8.000){2}{\rule{0.250pt}{0.400pt}}
\multiput(715.00,442.59)(0.494,0.488){13}{\rule{0.500pt}{0.117pt}}
\multiput(715.00,441.17)(6.962,8.000){2}{\rule{0.250pt}{0.400pt}}
\multiput(723.59,450.00)(0.485,0.569){11}{\rule{0.117pt}{0.557pt}}
\multiput(722.17,450.00)(7.000,6.844){2}{\rule{0.400pt}{0.279pt}}
\multiput(730.00,458.59)(0.494,0.488){13}{\rule{0.500pt}{0.117pt}}
\multiput(730.00,457.17)(6.962,8.000){2}{\rule{0.250pt}{0.400pt}}
\multiput(738.59,466.00)(0.488,0.560){13}{\rule{0.117pt}{0.550pt}}
\multiput(737.17,466.00)(8.000,7.858){2}{\rule{0.400pt}{0.275pt}}
\multiput(746.00,475.59)(0.494,0.488){13}{\rule{0.500pt}{0.117pt}}
\multiput(746.00,474.17)(6.962,8.000){2}{\rule{0.250pt}{0.400pt}}
\multiput(754.59,483.00)(0.488,0.560){13}{\rule{0.117pt}{0.550pt}}
\multiput(753.17,483.00)(8.000,7.858){2}{\rule{0.400pt}{0.275pt}}
\multiput(762.00,492.59)(0.494,0.488){13}{\rule{0.500pt}{0.117pt}}
\multiput(762.00,491.17)(6.962,8.000){2}{\rule{0.250pt}{0.400pt}}
\multiput(770.59,500.00)(0.488,0.560){13}{\rule{0.117pt}{0.550pt}}
\multiput(769.17,500.00)(8.000,7.858){2}{\rule{0.400pt}{0.275pt}}
\multiput(778.59,509.00)(0.488,0.560){13}{\rule{0.117pt}{0.550pt}}
\multiput(777.17,509.00)(8.000,7.858){2}{\rule{0.400pt}{0.275pt}}
\multiput(786.59,518.00)(0.488,0.560){13}{\rule{0.117pt}{0.550pt}}
\multiput(785.17,518.00)(8.000,7.858){2}{\rule{0.400pt}{0.275pt}}
\multiput(794.59,527.00)(0.488,0.560){13}{\rule{0.117pt}{0.550pt}}
\multiput(793.17,527.00)(8.000,7.858){2}{\rule{0.400pt}{0.275pt}}
\multiput(802.59,536.00)(0.488,0.560){13}{\rule{0.117pt}{0.550pt}}
\multiput(801.17,536.00)(8.000,7.858){2}{\rule{0.400pt}{0.275pt}}
\multiput(810.59,545.00)(0.485,0.645){11}{\rule{0.117pt}{0.614pt}}
\multiput(809.17,545.00)(7.000,7.725){2}{\rule{0.400pt}{0.307pt}}
\multiput(817.59,554.00)(0.488,0.626){13}{\rule{0.117pt}{0.600pt}}
\multiput(816.17,554.00)(8.000,8.755){2}{\rule{0.400pt}{0.300pt}}
\multiput(825.59,564.00)(0.488,0.560){13}{\rule{0.117pt}{0.550pt}}
\multiput(824.17,564.00)(8.000,7.858){2}{\rule{0.400pt}{0.275pt}}
\multiput(833.59,573.00)(0.488,0.626){13}{\rule{0.117pt}{0.600pt}}
\multiput(832.17,573.00)(8.000,8.755){2}{\rule{0.400pt}{0.300pt}}
\multiput(841.59,583.00)(0.488,0.560){13}{\rule{0.117pt}{0.550pt}}
\multiput(840.17,583.00)(8.000,7.858){2}{\rule{0.400pt}{0.275pt}}
\multiput(849.59,592.00)(0.488,0.626){13}{\rule{0.117pt}{0.600pt}}
\multiput(848.17,592.00)(8.000,8.755){2}{\rule{0.400pt}{0.300pt}}
\multiput(857.59,602.00)(0.488,0.626){13}{\rule{0.117pt}{0.600pt}}
\multiput(856.17,602.00)(8.000,8.755){2}{\rule{0.400pt}{0.300pt}}
\multiput(865.59,612.00)(0.488,0.626){13}{\rule{0.117pt}{0.600pt}}
\multiput(864.17,612.00)(8.000,8.755){2}{\rule{0.400pt}{0.300pt}}
\multiput(873.59,622.00)(0.488,0.626){13}{\rule{0.117pt}{0.600pt}}
\multiput(872.17,622.00)(8.000,8.755){2}{\rule{0.400pt}{0.300pt}}
\multiput(881.59,632.00)(0.488,0.626){13}{\rule{0.117pt}{0.600pt}}
\multiput(880.17,632.00)(8.000,8.755){2}{\rule{0.400pt}{0.300pt}}
\multiput(889.59,642.00)(0.488,0.626){13}{\rule{0.117pt}{0.600pt}}
\multiput(888.17,642.00)(8.000,8.755){2}{\rule{0.400pt}{0.300pt}}
\multiput(897.59,652.00)(0.485,0.798){11}{\rule{0.117pt}{0.729pt}}
\multiput(896.17,652.00)(7.000,9.488){2}{\rule{0.400pt}{0.364pt}}
\multiput(904.59,663.00)(0.488,0.626){13}{\rule{0.117pt}{0.600pt}}
\multiput(903.17,663.00)(8.000,8.755){2}{\rule{0.400pt}{0.300pt}}
\multiput(912.59,673.00)(0.488,0.692){13}{\rule{0.117pt}{0.650pt}}
\multiput(911.17,673.00)(8.000,9.651){2}{\rule{0.400pt}{0.325pt}}
\multiput(920.59,684.00)(0.488,0.692){13}{\rule{0.117pt}{0.650pt}}
\multiput(919.17,684.00)(8.000,9.651){2}{\rule{0.400pt}{0.325pt}}
\multiput(928.59,695.00)(0.488,0.626){13}{\rule{0.117pt}{0.600pt}}
\multiput(927.17,695.00)(8.000,8.755){2}{\rule{0.400pt}{0.300pt}}
\multiput(936.59,705.00)(0.488,0.692){13}{\rule{0.117pt}{0.650pt}}
\multiput(935.17,705.00)(8.000,9.651){2}{\rule{0.400pt}{0.325pt}}
\put(491,553){\makebox(0,0)[r]{Numerical}}
\put(161,129){\raisebox{-.8pt}{\makebox(0,0){$\circ$}}}
\put(231,151){\raisebox{-.8pt}{\makebox(0,0){$\circ$}}}
\put(301,160){\raisebox{-.8pt}{\makebox(0,0){$\circ$}}}
\put(371,198){\raisebox{-.8pt}{\makebox(0,0){$\circ$}}}
\put(441,224){\raisebox{-.8pt}{\makebox(0,0){$\circ$}}}
\put(511,277){\raisebox{-.8pt}{\makebox(0,0){$\circ$}}}
\put(580,319){\raisebox{-.8pt}{\makebox(0,0){$\circ$}}}
\put(650,388){\raisebox{-.8pt}{\makebox(0,0){$\circ$}}}
\put(720,446){\raisebox{-.8pt}{\makebox(0,0){$\circ$}}}
\put(790,532){\raisebox{-.8pt}{\makebox(0,0){$\circ$}}}
\put(860,606){\raisebox{-.8pt}{\makebox(0,0){$\circ$}}}
\put(930,708){\raisebox{-.8pt}{\makebox(0,0){$\circ$}}}
\put(561,553){\raisebox{-.8pt}{\makebox(0,0){$\circ$}}}
\end{picture}
\vskip 0.5cm
\parbox[t]{8 truecm}{\small
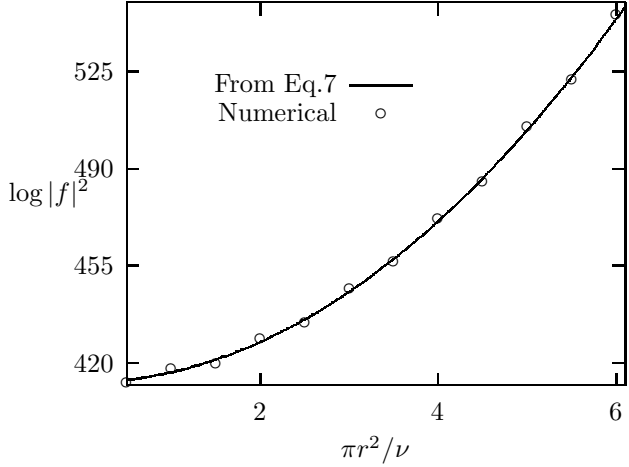
Figure 1: The log of the vortex function is plotted against $r^2$, both
from exact numerical computation and from Eq.7. The size of the lattice
is about R=6.3, containing about $\pi R^2/\nu \sim$ 145 vortex lattice points. 
}
\vskip 0.5 cm
\begin{equation}
|\Phi|^2=|c_q|^2({q\nu\over\pi})^q e^{-r^2\beta^2(1-\alpha)\over a_{xy}^2}, 
\alpha\equiv
{\pi a_{xy}^2\over \nu \beta^2}.
\end{equation}
In the above $\alpha$ is the ratio of the area of the effective
Larmor circle 
and the unit cell area of the vortex lattice.
In Fig.1 we plot the numerically computed function $\log |f|^2$, for a 
a lattice with 145 sites, along with the approximate value that
we calculated by doing an integral. 
As can be seen from the figure, the
above computation (Eq.7) is in excellent agreement with the numerical 
computation, implying the finite-size effects are quite small here.  

The total condensate density $|\psi|^2$ has now three variational parameters,
namely, $\alpha$ the number of vortices in an effective Larmor circle, 
$\beta$ the inverse rms width in the xy plane, $\gamma$ the inverse rms 
width along z direction.  The coefficients $\Gamma_1, \Gamma_2$
are calculated as
$$
\Gamma_1= {1\over 1-\alpha},\Gamma_2=1-\alpha.
$$
In the limit of $\alpha\rightarrow 0, \Gamma_1=\Gamma_2=1$, we have the
no vortex state.
Now, our task is to minimize the Gross-Pitaevskii energy (Eq.2) over variation 
of the parameters $\alpha,\beta,\gamma$. Setting the derivatives of $K(\alpha,
\beta,\gamma)$ w.r.t. the three variational parameters to zero, we obtain
\begin{eqnarray}
{1\over2}({1\over \beta^2}+\beta^2)-{\Omega\over \omega_{xy}}=
G\gamma \beta^2(1-\alpha)^2 \\
{1\over2} ({1\over \beta^2}-\beta^2)=
G\gamma \beta^2(1-\alpha)^2 \\
{\omega_z\over 2\omega_{xy}}({1\over \gamma^2}-\gamma^2)= G\gamma\beta^2(1-\alpha)
\end{eqnarray}
The first two equations imply $\beta^2=\Omega/\omega_{xy}$, and the last two
imply
$$
({1\over \gamma^2}-\gamma^2)^2= 2G\gamma(1-\beta^4){\omega_{xy}^2\over
\omega_z^2}.
$$
In the limit of $\Omega\rightarrow\omega_{xy}, \beta^2$ tends to one, and
$\gamma$ will also approach unity, as can be seen from the above equation.
However, even for $\Omega/\omega_{xy}=0.9$, the right hand side of the
equation is quite large (because $G=10^4,\omega_{xy}/\omega_z=4$), implying
$\gamma$ is quite small. Here, we can approximate ${1\over \gamma^2}-\gamma^2
\approx {1\over \gamma^2}$(this is tantamount to dropping the kinetic energy
term in $H_z$ (Eq.1), as usually done in the Thomas-Fermi approximation\cite
{ref2}), and thus obtain (valid for $\Omega>\Omega_c$)
$$1-\alpha={(1-\beta^4)^{3\over5}\over \beta^2(2G)^{2\over5}}({\omega_{xy}\over 
\omega_z})^{1\over5},
\beta^2={\Omega\over\omega_{xy}}$$,
$$\gamma=({\omega_z\over\omega_{xy}})^{2\over5}
{1\over (2G)^{1\over 5}(1-\beta^4)^{1\over5}}.
$$ 
As $\Omega\rightarrow \omega_{xy}$, $\beta^2\rightarrow 1,
\alpha\rightarrow 1$, which would imply $\gamma\rightarrow 1$, which
implies the rms width along z direction will tend to $a_z$. In the
neighborhood of $\Omega \le \omega_{xy}$, $\gamma$ has the form shown in the
above equation. Similarly, as $\Omega\rightarrow 0$, $\beta^2=\beta_0^2=0.16,
\gamma=\gamma_0=0.08,\alpha=0.$
The rms
widths of the condensate in the xy plane, and along z direction are
$$
R_{xy}=\sqrt{<r^2>}={a_{xy}\over \beta\sqrt {1-\alpha}},
R_z=\sqrt{<z^2>}={a_z\over \sqrt{2}\gamma}.
$$
In the quantum-Hall limit ($\Omega=\omega_{xy}$), the unit
cell area of the vortex lattice becomes $\nu =\pi a_{xy}^2$, implying as
many vortices as the number
of particles in the condensate.
The width of the condensate becomes very large in the
xy plane, and the width along z direction becomes very small (and saturates
at $a_z/\sqrt{2}$), as can be seen from the above relations. 

\vskip 0.7cm
\parbox[t]{8 truecm}{\small
Table 1: For various values of the rotational frequency, the number of
vortices within a size of the rms width of the condensate density, the
rms width of the condensate in the xy plane, and the rms width along
z direction. We have used the
experimental values of
$G=10^4$, and $\omega_z/\omega_{xy}=1/4$. Note that within a size of
$1.5R_{xy}$ (the condensate density falls to 10 per cent) the number of
vortices is 2.25 times that shown in the second column; for example the
number of vortices in the condensate is effectively about 25 for 
the second case shown.
As $\Omega$ increases the
rms width along z direction decreases, which will saturate at $a_z/\sqrt{2}$ 
in the limit $\Omega\rightarrow \omega_{xy}$. 
}
\vskip 0.5cm
\hglue 1cm
\begin{tabular}{|c|c|c|c|}
\hline
$\Omega/ \omega_{xy}$&$\pi R_{xy}^2/\nu$&$R_{xy}/a_{xy}$&$R_z/a_z$\\
\hline
  0   &  0  & 6.25 &  8.8\\
\hline 
  0.3 &  11 & 6.5  &  7.9\\
\hline
  0.5 &  22 &  6.9 &  7.5 \\
\hline
  0.7 &  40 &  7.7 &  6.9 \\
\hline
  0.9 &  96 & 10.4 &  5.7 \\
\hline
  0.94& 135 & 12.0 &  5.2 \\
\hline
  0.96& 175 & 13.5 &  4.7 \\
\hline
  0.98& 269 & 16.6 &  4.2 \\
\hline
\end{tabular}
\vskip 0.5cm
The aspect ratio $R_z/R_{xy}$ becomes very
small in this limit, and the system is essentially two-
dimensional, in analogy with the quantum-Hall system. The excitations of the
system are in xy plane. For $\Omega=\omega_{xy}$
we should go back to the original
problem of interacting Bose particles, rather than working within
a Gross-Pitaevskii paradigm, as small perturbations hitherto neglected will
become important. 

The optimal density of vortices can be written as
$$
{1\over\nu}= {\alpha\beta^2\over \pi a_{xy}^2}={2M \Omega\over h}\alpha
$$
which reduces to the familiar $1/\nu =2M\Omega/h$ in the limit of $\Omega
\rightarrow \omega_{xy}$. The optimal number of vortices within a size of
the rms width of the condensate is
$$
{\pi <r^2>\over \nu}= {\alpha\over 1-\alpha}.
$$
The optimal Gross-Pitaevskii energy is given by
$$
K={5\over4}{2^{1\over5}\over \pi^{1\over5}}N^{7\over5} ({\omega_z\over \omega_{
xy}})^{1\over5} ({a_{sc}\over a_z})^{2\over5}(1-{\Omega^2\over \omega_{xy}^2})
^{2\over 5} \hbar \omega_{xy}.$$
In Table 1 we have listed the optimal values of the variational parameters,
the rms widths, the number of vortices within the rms width, for various
values of the rotational frequencies. These results are different from
the work of \cite{ref2}.
For $\Omega/\omega_{xy}\sim 0.9$, there are a large number of vortices, and
$\gamma$ is still small, the density profile along z direction is similar to
the inverted parabola of\cite{ref2}, except near the edge where it has a 
gaussian tail. But, as $\Omega$ becomes closer to 
$\omega_{xy}$, as $\gamma\rightarrow 1$, the density profile along z direction
is that of the harmonic oscillator groundstate (the rms width is equal to 
$a_z$). A quantitative measure of how far the system is from quantum-Hall 
limit can be taken as $1-\gamma$.

In the foregoing analysis, the leading order behaviour is valid for $\Omega\sim
\omega_{xy}$, where a large number of vortices are generated. The lattice
structure of the vortices enters only through the density. The triangular
lattice is favored over the square lattice, as for the same density of vortices
($\nu=\sqrt{3}c_{tr}^2/2=c_{sq}^2$, where $c_{tr}, c_{sq}$ are the lattice
spacing for the triangular and square lattices respectively), the spacing is
larger in the case of triangular lattice. Since the vortices have an effective
repulsion, a larger spacing is favored for energy minimization.
But, in the quantum-Hall regime ($\Omega=\omega_{xy}$), the energies of 
the two lattice
structures become degenerate (square lattice energy approaches that of the
triangular lattice from above). Then it becomes important to consider all
perturbations, though small and hence hitherto neglected, to lift the degeneracy
to find the true groundstate configuration. This implies one should go beyond
the Gross-Pitaevskii for the condensate wavefunction. There is also a 
possibility of
formation of higher angular momentum vortices (for $\Omega>\omega_{xy}$), and
maybe an instability of fragmenting of the condensate. This is because, once
the unit cell area becomes equal to the area of Larmor circle, the angular
momentum that can be carried by $m=1$ vortices gets saturated. If a larger 
angular momentum is
transfered to the system, $m>1$ get generated, or the system breaks into
smaller pieces. It will be interesting to investigate other variational states
in this regime, as the physics here is similar to that of a Quantum Hall system.
We are investigating these issues currently, along with the question of
multi-component Bose condensates where the internal degree of freedom introduces
many features motivating many novel variational states.

In conclusion, we have calculated the critical rotational frequency $\Omega_c/
\omega_{xy}\sim 0.07$, above which vortices become stable, which compares
very well with the MIT experiments. We have shown that variational states,
with a gaussian density profile along z direction, and density profile with
vortices of unit strength , the locations forming a triangular lattice, 
in the xy plane get stabilized for larger $\Omega$. We have computed the width
of the wavefunctions along z direction, and in the xy plane, the aspect
ration becoming very small as $\Omega\rightarrow \omega_{xy}$. The unit cell
area of the vortex lattice becomes equal to the
Larmor circle area in this limit, implying as many vortices as the number of
condensate atoms. 

It is a great pleasure to thank Prof. V. Ravishankar for extensive discussions
and a critical reading of the manuscript.

\vskip 0.4 cm
\end{document}